\documentclass[conference,letterpaper]{IEEEtran}
\IEEEoverridecommandlockouts
\usepackage{cite}
\usepackage{amsmath,amssymb,amsfonts}
\usepackage{algorithmic}
\usepackage{graphicx}
\usepackage{textcomp}
\usepackage{xcolor}
\usepackage{url}
\usepackage{balance}

\usepackage[acronym]{glossaries}
\usepackage{glossaries-prefix}
\usepackage{url}
\usepackage{tabularx}
\usepackage{multirow}
\usepackage{subfig}
\usepackage{booktabs}
\usepackage{enumitem}

\definecolor{red3}{RGB}{136,0,3}
\definecolor{comment}{named}{red3}

\newboolean{showcomments}
\setboolean{showcomments}{true} %
\ifthenelse{\boolean{showcomments}}
  {
		\newcommand{\nbb}[2]{
		\fcolorbox{black}{yellow}{\bfseries\sffamily\scriptsize#1}
		{\sf$\blacktriangleright$\textcolor{blue}{\textit{#2}}$\blacktriangleleft$}
		}
		
		\newcommand{\remarks}[1]{\color{red}[#1]\color{black}}

		\newcommand{\del}[1]{\textcolor{red}{\sout{#1}}} %
  }
  {
		\newcommand{\nbb}[2]{}
		\newcommand{\remarks}[1]{}

		\newcommand{\del}[1]{} %
  }

\newacronym{AI}{AI}{Artificial Intelligence}
\newacronym{AEB}{AEB}{Autonomous Emergency Braking}
\newacronym[%
    longplural={Automated Driving Systems},
    shortplural={ADS}
    ]
    {ADS}{ADS}{Automated Driving System}
\newacronym[%
    longplural={Advanced Driver Assistance Systems},
    shortplural={ADAS}
    ]
    {ADAS}{ADAS}{Advanced Driver Assistance System}
\newacronym[%
    longplural={Advanced Driver Assistance Systems / Automated Driving},
    shortplural={ADASs/AD}
    ]
    {ADAS/AD}{ADAS/AD}{Advanced Driver Assistance System / Automated Driving}
\newacronym{DRM}{DRM}{Design Research Methodology}
\newacronym{DSR}{DSR}{Design Science Research}
\newacronym{LDW}{LDW}{Lane Departure Warning}
\newacronym{ML}{ML}{Machine Learning}
\newacronym{ODD}{ODD}{Operational Design Domain}
\newacronym[%
    longplural={Original Equipment Manufacturers},
    shortplural={OEMs}
    ]
    {OEM}{OEM}{Original Equipment Manufacturer}
\newacronym[
    prefixfirst={a\ },%
    prefix={an\ }%
    ]
    {SFC}{SFC}{Space-Filling Curve}
\newacronym{SOTIF}{SOTIF}{Safety of the intended functionality}
\newacronym{SUT}{SUT}{System Under Test}
\newacronym{VR}{VR}{virtual reality}
\newacronym{ZEBRA}{ZEBRA}{Z-order Curve-based Event Retrieval Approach}

\def\BibTeX{{\rm B\kern-.05em{\sc i\kern-.025em b}\kern-.08em
    T\kern-.1667em\lower.7ex\hbox{E}\kern-.125emX}}
\begin{document}

\title{
ZEBRA: Z-order Curve-based Event Retrieval Approach to Efficiently Explore Automotive Data
}

\author{\IEEEauthorblockN{Christian Berger}
\IEEEauthorblockA{
\textit{University of Gothenburg}\\
Gothenburg, Sweden \\
christian.berger@gu.se}
\and
\IEEEauthorblockN{Lukas Birkemeyer}
\IEEEauthorblockA{
\textit{Technische Universität Braunschweig}\\
Braunschweig, Germany \\
l.birkemeyer@tu-braunschweig.de}
}

\maketitle

\begin{abstract}
Evaluating the performance of software for automated vehicles is predominantly driven by
data collected from the real world. While professional test drivers are supported with
technical means to semi-automatically annotate driving maneuvers to allow better event
identification, simple data loggers in large vehicle fleets typically lack automatic and
detailed event classification and hence, extra effort is needed when post-processing such
data. Yet, the data quality from professional test drivers is apparently higher than the
one from large fleets where labels are missing, but the non-annotated data set from large
vehicle fleets is much more representative for typical, realistic driving scenarios to be
handled by automated vehicles. However, while growing the data from large fleets is
relatively simple, adding valuable annotations during post-processing has become increasingly
expensive. In this paper, we leverage Z-order space-filling curves to systematically reduce
data dimensionality while preserving domain-specific data properties, which allows us to
explore even large-scale field data sets to spot interesting events orders of magnitude faster
than processing time-series data directly. Furthermore, the proposed concept is based on an
analytical approach, which preserves explainability for the identified events.
\end{abstract}


\section{Introduction}
Since \glspl{ADAS} and \glspl{ADS} of modern vehicles highly interact with their environment,
researchers and engineers face challenges in handling the vast space of possible scenarios during
the development and testing of such systems. To better cope with the sheer unbounded unstructuredness
during the development phase, researchers and engineers are exploring the proper use of
\gls{AI}/\gls{ML}-based components to improve perception systems, for example, for pedestrian detection~\cite{ZBD21}. The performance of these components, though, depends on the quality and
diversity of the data, which was used to train such \gls{ML}-based components. Moreover, the data
set shall ideally reflect relevant and important aspects of the intended \gls{ODD}. Hence, start-up
companies as well as legacy automotive \glspl{OEM} began in recent years to collect enormous amounts
of data to systematically train and test \gls{ADAS}/\gls{ADS}. This growing amount of data is partially
annotated, fueling virtual environments for closed-loop testing, or to infer information about
\emph{typical} traffic scenarios to deal with on average.

\subsection{Problem Domain \& Motivation}
\label{sec:research-problem}
Data quality and data diversity are primary drivers to determine and improve the performance of
\gls{AI}/\gls{ML}-based software. However, just growing existing data sets into barely manageable
data lakes is an insufficient response to meet quality criteria for a safer engineering and testing of
\gls{AI}/\gls{ML}-based software. Instead, large-scale properly annotated data sets or efficient
filtering methods are required to support event queries such as ``retrieve data samples from a
specific time of the year, in a specific region, and containing certain traffic participants that
perform a lane change''. Annotating data to realize this query is an expensive operation that
demands enormous effort and is potentially prone to biases. A more efficient way to retrieve events
from data sets is to initially annotate data samples automatically, which, however, requires
knowledge of relevant clusters. Updating the set of relevant clusters to include new participants
such as e-scooters for example may cause an expensive re-annotation of all existing entries in a data set. 

Alternatively, filtering the data based on predefined search criteria using selected data properties
is an automated operation that reduces human impact. However, na\"ively evaluating all entries of
large data sets entry by entry does not scale. As state-of-the-art configurations for automated vehicles
generate between 1-3TB of data per
hour\footnote{Cf.~\url{https://developer.nvidia.com/blog/training-self-driving-vehicles-challenge-scale/}},
all previously mentioned approaches are inappropriate due to the enormous effort spent for annotation or
because of computational inefficiency when filtering such time-series data. Hence, we motivate our work
as follows:
\\

\noindent\fbox{%
\begin{minipage}{.96\columnwidth}
\emph{Motivation:} Computationally efficient approaches are needed to systematically describe \emph{and}
analytically understand the inherent value of growing data lakes to better support the data-driven development
and improvement of \gls{AI}/\gls{ML}-based automotive software.
\end{minipage}
}\\

\subsection{Research Goal \& Research Questions}

We pursue the research goal to design and evaluate an approach to computationally efficiently
identify and retrieve events from large-scale automotive data sets. In this paper,
we leverage the dimensionality reducing property of \glspl{SFC} to traceably compute a
domain-specific, compact representation that we exploit as a novel index to entries of a
large-scale data set for retrieving events based on efficient range queries.
We conducted experiments to answer the following research questions:

\noindent\fbox{%
\begin{minipage}{.96\columnwidth}
\begin{description}[style=unboxed,leftmargin=0cm]
    \item[RQ-1:] How can \glspl{SFC} be exploited to retrieve automotive events?
    \item[RQ-2:] What is the performance of event retrieval based on \glspl{SFC}?
\end{description}
\end{minipage}
}\\

\subsection{Contributions \& Scope}
In this paper, we leverage \glspl{SFC} to provide the \gls{ZEBRA} to efficiently process large-scale
data sets. We evaluate the efficacy and efficiency of \gls{ZEBRA} for detecting maneuvers based on
multi-dimensional acceleration data on an automotive data set. Yet, its underlying concept and
implementation are generic and designed to handle various, multi-dimensional, numerical data. 

\subsection{Structure of the Article}
The remaining part of the paper is structured as follows: Sec.~\ref{sec:background} introduces
briefly the concept of \glspl{SFC} to provide the relevant terminology and theoretical concepts.
In Sec.~\ref{sec:relatedworks}, we present and discuss related works. We introduce \gls{ZEBRA}
in Sec.~\ref{sec:ZEBRA} and discuss results from a systematic evaluation in Sec.~\ref{sec:evaluation}
before we conclude our work in Sec.~\ref{sec:conclusion}.

\section{Background}
\label{sec:background}

\begin{figure}[t!]
    \centering
    \includegraphics[trim=10 0 0 10, clip,width=\linewidth]{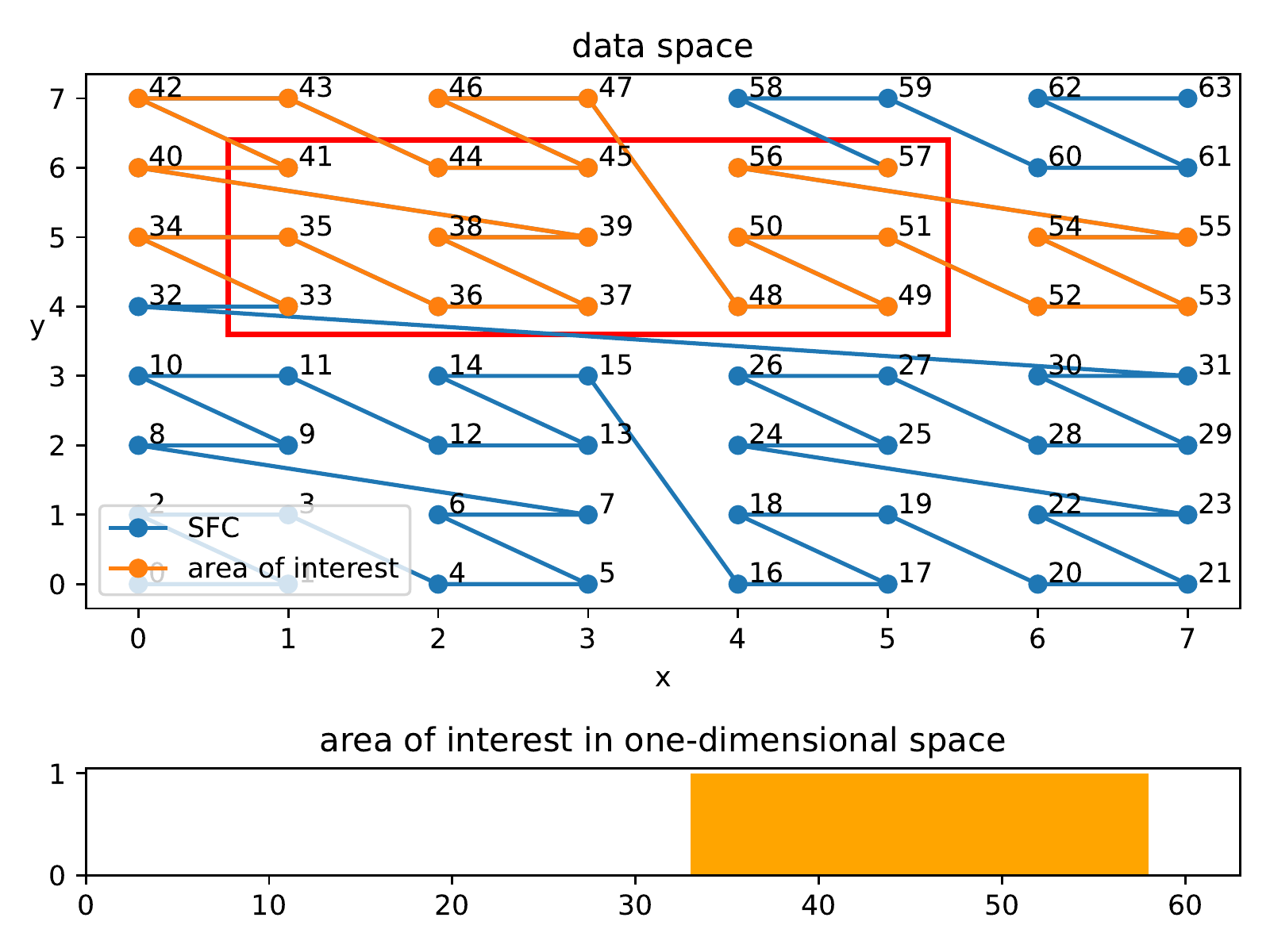}
    \caption{
    Application of a \gls{SFC} (blue line) to reduce dimensions from a two-dimensional space (above) to a single-dimensional space (below). An arbitrary area of interest indicated by a red rectangle in the two-dimensional space selects a sub-set of the data (orange) that is highlighted in both spaces.
    }
    \label{fig:locality_dataspace}
\end{figure}

\glspl{SFC} are used to map multi-dimensional values to their corresponding \emph{single-dimensional}
representations (cf.~Bader, \cite{Bad13}), which can be exploited for computationally efficient data
exploration (cf. Fig.~\ref{fig:locality_dataspace}). An early definition of \glspl{SFC} can be found by
Goldschlager who states that ``a space-filling curve is a continuous curve [\dots], which passes through
every point in the square'' \cite{Gol81}. This points-passing curve is constructed following a very
specific repetitive or recursive pattern to pervade a given space.

In Fig.~\ref{fig:locality_dataspace}, we present an example of the repeating Z-order curve
(blue zig-zag-line) covering the complete two-dimensional value space. For this purpose, the shape of
the \gls{SFC} is determined by a dimensionality reducing mapping function like Morton-codes
(cf.~Bader, \cite{Bad13} and Morton, \cite{Mor66}) that associates a single-dimensional value to every
$n$-tuple from the $n$-dimensional data space. The mathematical operation to turn a multi-dimensional
data sample into its one-dimensional representation when using a Z-order curve is \emph{bit-interleaving}.
This transformation is traceable, explainable, and bi-directional.

Querying the multi-dimensional value space is depicted as a red rectangle in Fig.~\ref{fig:locality_dataspace}
denoting an area-of-interest, where relevant points on the single-dimensional Z-order curve are
highlighted in orange. In the single-dimensional space, the Morton-codes for the bottom/left corner
($x=1, y=4$) and top/right corner ($x=5, y=6$) are computed as $33_{\text{Morton}}$ and $57_{\text{Morton}}$,
respectively. The complete value range between these two Morton-codes does not contain any false/negatives,
ie., multi-dimensional values that should have been included in the area-of-interest but were not. Moreover,
\glspl{SFC} exhibits \emph{locality preservation} (cf.~Bader, \cite{Bad13}), which illustratively means
that points that are in close proximity in the multi-dimensional space will also be located near to each
other on a \gls{SFC}. For example, \glspl{SFC} found an application for looking up data samples from a database that reside
in specific geographical regions (cf.~Sec.~\ref{sec:relatedworks}).

\section{Related Works} %
\label{sec:relatedworks}

A na\"ive approach to retrieve events from large data sets is brute-force that simply traverses all
entries to find matching properties. Perez et al.~\cite{PEREZ201710} suggested an approach based on such
an idea to use kinematics thresholds for event detection. While event retrieval is rule-based and traceable,
its actual effort scales with the number of entries in a data set and hence, brute force
is practically infeasible to query events from large-scale data sets.

Improved approaches use clustering techniques for event detection by analyzing patterns with respect
to their similarity and classify them into clusters.
Current approaches use for example the k-means algorithm as suggested by Hauer et al.~\cite{hauer2020clustering}
or \gls{ML}-enabled approaches such as auto-encoders as proposed by Demetriou et al., Hoseini et al., and
Ding et al.~\cite{demetriou2020deep, hoseini2021vehicle, ding2019multi} to cluster entries from a data
set. Initially determining existing clusters and assigning relevant entries from a data set to these clusters
require significant computational effort. Inserting and assigning single entries retroactively, though,
is a computationally efficient operation. However, such clusters need to be updated regularly as the
characteristics of data set entries might evolve over time.

Another research direction as suggested by Hulbert et al.~\cite{HKH+16} suggests to use \gls{SFC} to
realize data sample look-ups in large-scale data sets. Bader (cf.~\cite{Bad13}) and Morton (cf.~\cite{Mor66})
for example suggest to use Z-order curve to turn a tuple of (latitude, longitude) GPS coordinates into
a single-dimensional representation allowing a simple retrieval of all GPS coordinates of interest by looking
up their corresponding Morton-encoded, single-dimensional representations within an area of interest thanks
to the \gls{SFC}'s locality preservation property.
Inserting new entries into a data set, generating
a \gls{SFC}-index, and retrieving entries from an area of interest are computationally efficient operations.
Moreover, calculating a \gls{SFC}-index for related entries in a data set is a rule-based, traceable operation.

We are proposing in this work the combination of the locality preservation property of \glspl{SFC} with temporal
properties of consecutive data samples from \gls{ODD} scenarios such as automotive data as a novel approach
to model and retrieve events of interest.

\section{ZEBRA: Z-order Curve-based\\ Event Retrieval Approach}
\label{sec:ZEBRA}

\subsection{Overview of the Fundamental Concept}
\begin{figure}[h!]\centering
    \includegraphics[width=\linewidth]{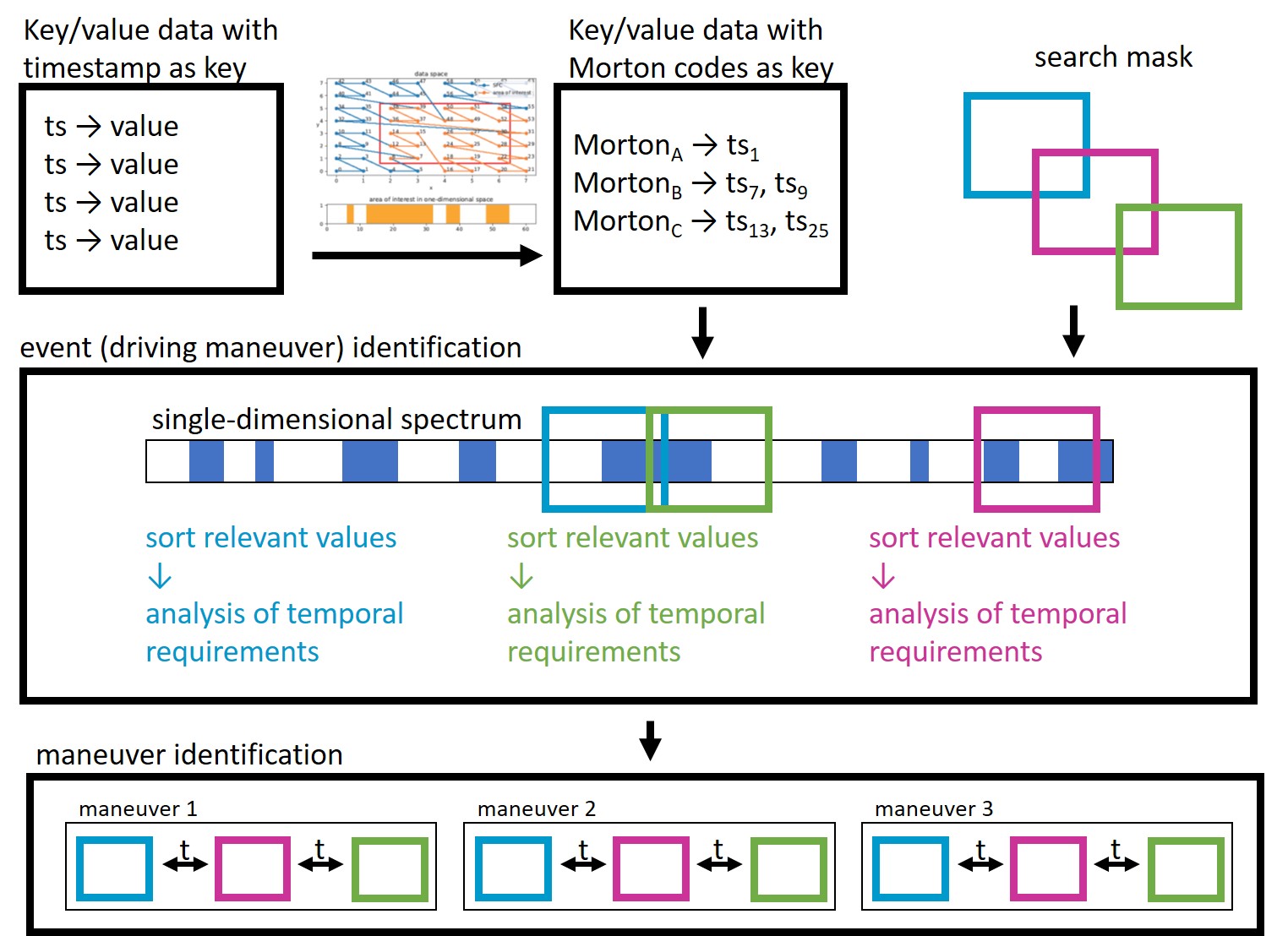}
    \caption{Overview of the approach \gls{ZEBRA} to model and retrieve events on the example of automotive maneuver identification.}
    \label{fig:architecture}
\end{figure}

The fundamental idea behind our proposed \emph{Z-order Curve-based Event Retrieval Approach} (\gls{ZEBRA})
exploits the dimensionality reduction from \glspl{SFC} to transform multi-dimensional data samples into
a single-dimensional representation in combination with the temporal occurrence of such data samples
as depicted in Fig.~\ref{fig:architecture}. The visual representation of the resulting single-dimensional
data resembles a \emph{Zebra-pattern} as depicted in Fig.~\ref{fig:braking_single_round}(c). The specific
ordering, temporal occurrence and spread, and distribution of the resulting stripes correspond semantically
to scenarios in the \gls{ODD} that we refer to as events; in the automotive \gls{ODD} case as described in
Sec.~\ref{sec:evaluation}, such events correspond to driving maneuvers. In \gls{ZEBRA}, we are now using
the resulting single-dimensional representations as additional index for an existing key/value database where
all data samples are stored. This index, which contains domain-specific properties from the original
multi-dimensional space, is ordered and thus, can be computationally efficiently exploited for range queries
to retrieve events of interest. We use search masks (ie., area of interests) to query individual events and
multi-level search masks to query for temporal combinations of individual events. Thus, we are
able to retrieve arbitrary events that are defined by search masks from large-scale databases by exploiting
the temporal relations between \emph{Zebra-pattern}-like stripes.

\subsection{Application of \gls{ZEBRA} to Automotive \glspl{ODD}}

In the following, we outline how we use this approach for event retrieval on automotive data sets.
We consider vehicle kinematics over time as multi-dimensional value space $\mathbb{R}^n$ for event
modeling in automotive data sets. Individual vectors contain the vehicle's acceleration or steering wheel
angle for example. Using \gls{ZEBRA}, these vectors are discretized to $\mathbb{Z}_+^n$ to allow for
applying dimensionality reduction using Morton-codes for the Z-order curve. The obtained Z-order curve
represents only actually existing points in the multi-dimensional value space as
\emph{Zebra}-stripes in their corresponding, single-dimensional representation. Values that do not occur
in the multi-dimensional data space also do not occur in the single-dimensional \gls{SFC} representation. 

Next, the specific ordering, temporal occurrence and spread, and distribution of these stripes in the
single-dimensional representation need to be interpreted within the originating \gls{ODD} of the automotive
data space for event modeling and retrieval. For this step, the temporal dimension of the \gls{ODD} is
considered that intuitively determines what stripes occur at what point in time and what their temporal
distance and spread is. Different scenarios from the \gls{ODD} exhibit different yet specific characteristics
for stripe ordering, temporal occurrence and spread of stripes, and stripe distribution that we exploit
in our search masks to retrieve events from a data set. In Sec.~\ref{sec:evaluation}, we are providing
two examples from the automotive \gls{ODD}: \gls{AEB} and \gls{LDW}, on which we apply and evaluate \gls{ZEBRA}.

\section{Evaluation of \gls{ZEBRA} and Discussion}
\label{sec:evaluation}

In this section, we describe our experimental results from applying and evaluating \gls{ZEBRA} to
efficiently retrieve events from automotive data to address the aforementioned research questions.

\subsection{Experimental Setup}
To evaluate how \gls{ZEBRA} can be used to retrieve events from automotive \gls{ODD}, we set up an experiment
on a confined test site where we collected data using a Volvo XC90 instrumented with an Applanix GPS/IMU system.
We defined two driving maneuvers: (a) Emergency braking as found in \gls{AEB} systems for example, and (b) lane
change maneuvers as found in \gls{LDW} systems. For that purpose, we drove the vehicle manually and conducted
the required maneuvers at different speeds while recording accelerations, angular velocities, and the
steering wheel angle. We used this controlled experiment to evaluate whether \gls{ZEBRA} can correctly identify
these maneuvers. Afterwards, we collected data during a test drive through Gothenburg in Sweden where we also
included lane changes that we manually annotated during the data collection with the goal to let \gls{ZEBRA}
spot these events automatically afterwards. Due to safety reasons, we excluded emergency braking scenarios
during the real-world data collection in Gothenburg. For all collected data sets, we generated \gls{SFC} representations
and use \gls{ZEBRA} to retrieve emergency braking and lane change
events that we compared afterwards with the manual annotations.

\begin{figure}[t!]%
    \centering
    \subfloat[][Vehicle position during two emergency braking maneuvers conducted on a confined test area: The vehicle is driving an oval, accelerating, and conducting twice a harsh braking maneuver (highlighted in red).]{%
        \includegraphics[trim=107 390 20 25, clip,width=\linewidth]{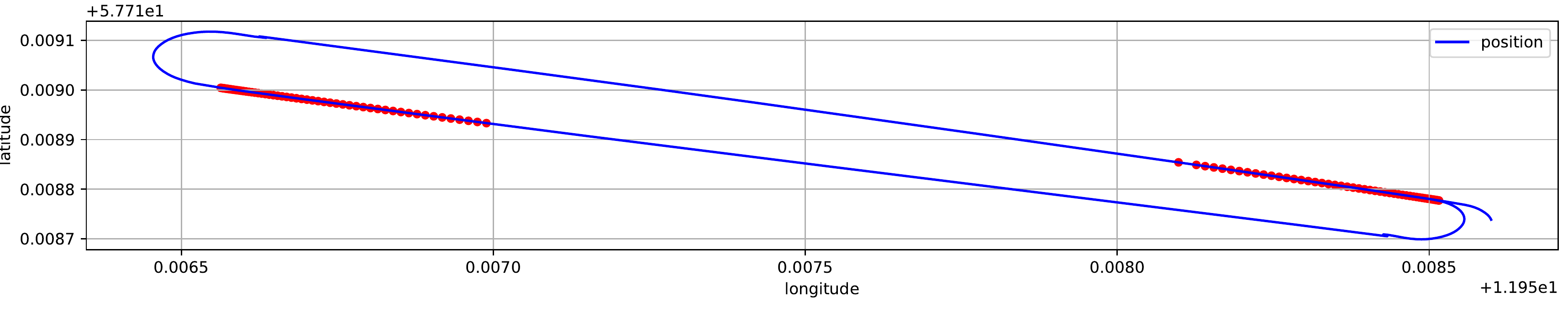}
    }%
    \qquad
    \subfloat[][Vehicle's accelerations for the X-Y-dimensions over time: The red \emph{hats} laid over the x-acceleration depict the two emergency braking maneuvers.]{%
        \includegraphics[trim=107 170 20 260, clip,width=\linewidth]{pics/braking_eine_runde.pdf}
    }%
    \qquad
    \subfloat[][\emph{Zebra}-pattern representing the single-dimensional Morton codes corresponding to the acceleration tuples from Fig.~(b)
    ]{%
        \includegraphics[trim=110 40 20 480, clip,width=\linewidth]{pics/braking_eine_runde.pdf}
    }%
    \caption{Illustrating example for using \gls{ZEBRA} to spot emergency braking scenarios
    }
    \label{fig:braking_single_round}
\end{figure}%

\subsection{Using \gls{ZEBRA} for Detecting Emergency Braking Scenarios}
The first application of \gls{ZEBRA} is depicted in Fig.~\ref{fig:braking_single_round}, where \gls{ZEBRA}
is used to spot emergency braking scenarios in a data set recorded on a confined test site.
The vehicle was accelerated to inner city driving speed before an emergency braking maneuver was conducted.

When processing these data samples using a brute force approach, an event detection algorithm would have to
process all rows using a sliding window for example to find rising edges over time for the longitudinal
acceleration of the vehicle (cf.~Fig.~\ref{fig:braking_single_round}(b)) resulting in a computational complexity
of at least $O(n)$. When operating on a Z-order curve, though, relevant events can simply be spotted by using a
threshold query exhibiting better than $O(n)$ performance as all tuples with excessive longitudinal/lateral-accelerations will be located in a certain part
of the single dimensional spectrum after transforming such multi-dimensional data to their single-dimensional
representations as depicted using red stripes in Fig.~\ref{fig:braking_single_round}(c): They are all located
at the right hand side. As illustrated in Sec.~\ref{sec:ZEBRA}, we use the single-dimensional representations
as additional index into a key/value database, which allows us to efficiently look-up all time points where
such emergency braking maneuvers were conducted.

\subsection{Using \gls{ZEBRA} for Detecting Lane Changes}
\begin{figure}[t!]%
    \centering
    \subfloat[][Vehicle position during two lane change maneuvers conducted on a confined test area: The vehicle is driving an oval, accelerating, and conducting twice a lane change maneuver from left to right lane (highlighted in red).]{%
        \includegraphics[trim=107 390 20 25, clip,width=\linewidth]{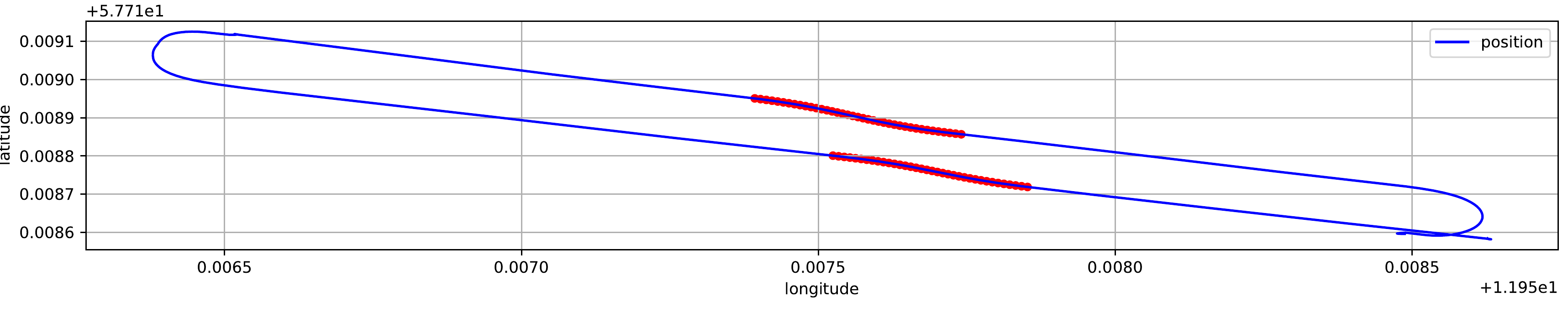}
    }%
    \qquad
    \subfloat[][Vehicle's accelerations for the X-Y-dimensions over time: The red rectangles laid over the y-acceleration depict the two lane change maneuvers.]{%
        \includegraphics[trim=107 170 20 260, clip,width=\linewidth]{pics/LC.pdf}
    }%
    \qquad
    \subfloat[][\emph{Zebra}-pattern representing the single-dimensional Morton codes corresponding to the acceleration tuples from Fig.~(b)]
    {%
        \includegraphics[trim=107 40 20 480, clip,width=\linewidth]{pics/LC.pdf}
    }%
    \caption{Illustrating example for using \gls{ZEBRA} to spot lane change scenarios}
    \label{fig:LC}
\end{figure}

The second application of \gls{ZEBRA} is shown in Fig.~\ref{fig:LC}, where \gls{ZEBRA} is used to identify
lane change maneuvers. In contrast to the previous example, where already a simple threshold query in the
single-dimensional space was sufficient, detecting lane changes may require the observation of a certain
acceleration profile in two dimensions over time. Hence, both aspects from \gls{ZEBRA}, dimensionality
reduction \emph{and} temporal properties from the \gls{ODD} need to be exploited to identify this type of
events. For this experiment, the same instrumented vehicle was used on a confined test site. The vehicle was
accelerated to inner city driving speed before a lane change maneuver from the left to the right lane was
conducted. In a second step, the same driving profile was repeated but the lane change maneuver was omitted.

Both runs were subsequently converted to a single-dimensional spectrum and the second run was subtracted
from the first run to identify relevant sections on the \gls{SFC}. These sections were then analyzed
to identify relations between occurring time points and spread of stripes over time as shown in
Fig.~\ref{fig:LC}(c) to model a search mask for an event detector. The design of this event detector contained
several stages as first, a range query was needed to spot the \emph{falling V-shapes} of candidates
for potential lane change maneuvers as highlighted with red rectangles in Fig.~\ref{fig:LC}(b). As the
corresponding Morton-code representations are used as additional index into a key/value database referring to time
points, such retrieved time points were stored in a list as input for the second stage of this event detector
that was spotting the second half of the \emph{flipped V-shaped} curve as highlighted in Fig.~\ref{fig:LC}(b);
the corresponding time points were then added to a second list. The final stage of this event detector was then
matching starting time points from the first list with time points from the second list that in combination had
to match semantic plausibility checks from the \gls{ODD} like minimal and maximal durations to be classified as
a lane change maneuver.

Similar to the previous example, traversing all rows using a sliding window with a variable look-ahead
to spot such scenarios would require at least $O(n)$. When operating on the Z-order curve, though, spotting
the two V-shapes can be realized as computationally efficient range queries exhibiting better than $O(n)$ performance and only the final stage in the
search mask of the event detector would require a linear traversal of potential candidate entries from both
lists, which is now only a small fraction from the entire data set.

\subsection{Using \gls{ZEBRA} on a Real World Example}

\begin{figure}[t!]%
    \centering
    \subfloat[][Vehicle position during the data collection highlighting lane changes to the left lane in red.]{%
        \includegraphics[trim=98 400 20 5, clip,width=\linewidth]{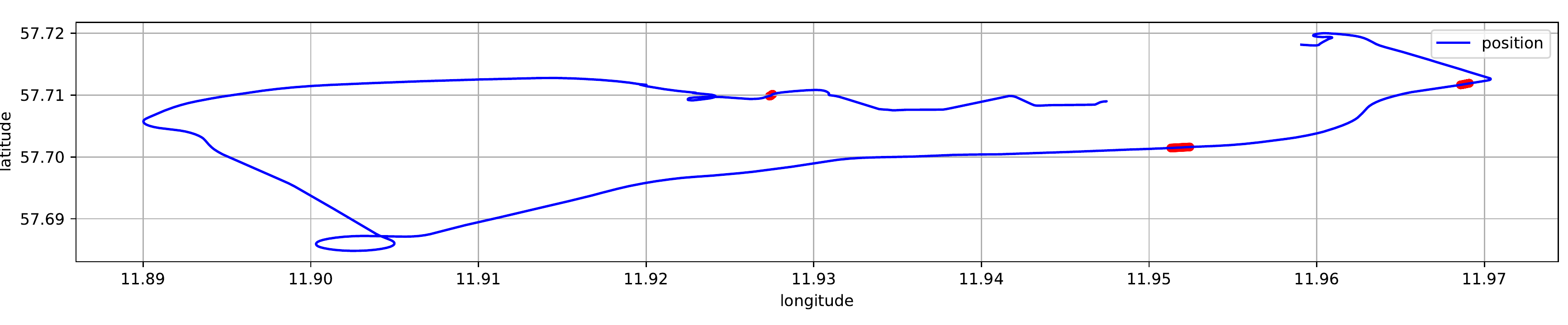}
    }%
    \qquad
    \subfloat[][Vehicle's accelerations for the X-Y-dimensions over time: Three red rectangle laid over the y-acceleration depict the lane change scenarios.]{%
        \includegraphics[trim=98 160 20 240, clip,width=\linewidth]{pics/Goeteborg_LC_to_left.pdf}
    }%
    \qquad
    \subfloat[][\emph{Zebra}-pattern representing the single-dimensional Morton codes corresponding to the acceleration tuples from Fig.~(b) on a Z-order curve.]{%
        \includegraphics[trim=107 40 20 480, clip,width=\linewidth]{pics/Goeteborg_LC_to_left.pdf}
    }%
    \caption{Illustrating example for using \gls{ZEBRA} to spot lane changes from the left to the right lane during a data collection in Gothenburg, Sweden.}
    \label{fig:Goeteborg_LC_to_left}
\end{figure}%

Finally, we conducted a data collection in Gothenburg, Sweden to evaluate the transferability of the event
detection approach calibrated on one data set recorded at a confined test site to a data set, on which the
event detector was not applied before. The same instrumented vehicle was used to collect GPS and acceleration
data for approx.~23 minutes as depicted in Fig.~\ref{fig:Goeteborg_LC_to_left}. All lane changes from right
to left lane and vice versa were manually annotated to be able to verify the results from \gls{ZEBRA} afterwards.
The recorded data was transformed from its multi-dimensional space into the single-dimensional spectrum and
the lane change event detector was applied thereto. During the test drive, three lane change maneuvers on
straight roads were conducted that were also spotted from \gls{ZEBRA}. Since we defined and calibrated the
search mask to detect lane changes for straight roads on flat surfaces, we only detect this type of lane changes
within our experiments. As the definition and detection of search masks, however, is generic, defining search
masks that cover lane changes when passing curves is possible by adjusting the search mask calibration parameters,
which was outside the scope of our experiments.
One observation can be made when comparing Fig.~\ref{fig:LC}(c) with Fig.~\ref{fig:Goeteborg_LC_to_left}(c):
The spectra are not exactly the same and this is motivated by the fact that the data collection during
driving through the city created obviously more tuples of longitudinal/lateral-accelerations than what was
possible with the controlled experiments on the confined test site only. Hence, we see similar sections in both
spectra covered by stripes but in addition, the spectrum from the larger test drive shows also broader ranges
covered by stripes.

\noindent\fbox{%
\begin{minipage}{.96\columnwidth}
\textbf{Answering RQ-1: How can \glspl{SFC} be exploited to retrieve automotive events?}\\
The \acrfull{ZEBRA} uses \glspl{SFC} to generate a domain-specific representation that we use as index into a
key/value database to model and computationally efficiently retrieve automotive events from annotated driving experiments
on a confined test sites and from a real-world data collection. We created \gls{SFC} representations for
tuples of longitudinal/lateral-accelerations and successfully retrieved emergency braking and lane changes maneuvers.
\end{minipage}
}\\

\subsection{Performance of ZEBRA}  %
\label{sec:results-discussion}

Next, we evaluate the performance of \gls{ZEBRA} by conducting an experiment that investigates its efficiency
and efficacy. We query maneuvers from a large-scale, real world data set collected in Gothenburg, Sweden
(cf.~Berger \cite{BKGB22}) with three event retrieval approaches: A) BF\_primitive, which is a na\"{i}ve
implementation processing the data set entry-by-entry from beginning to its end, ie., operating in
\emph{brute force} (BF) and hence, exhibiting $O(n)$; B) an improved version of the previous event detector
that scales better in case when multi-stage data look-ups are necessary;
and C) \gls{ZEBRA}: our \gls{SFC}-based implementation of the event detector.

The BF\_primitive query requires a key/value database with time points as keys and a search mask as input.
The output is a set of tuples of time points describing the start and end of detected driving maneuvers.
BF\_primitive traverses the entire database for each stage in the search mask and identifies relevant entries
(ie., whether longitudinal- and lateral-accelerations are in the defined value range). We analyze the time
points of relevant values according to their temporal relations and consider them as a valid driving maneuver
when they match a minimum and a maximum duration. However, we allow outliers as long as the temporal gap between
two following relevant values does not exceed a defined threshold.
In case of a multi-stage search mask, we also investigate the temporal gap between two following stages and
define a minimum and a maximum gap.

The improved BF query also requires a key/value database with time points as keys and it outputs a combination of
start and end time points that describe valid maneuvers. The improved BF query iterates the entire database only
once and determines entries that are valid according to the first search mask stage (ie., value range and temporal
relations). After a valid starting part of a driving maneuver has been detected according to the first search mask
stage, the improved BF query recursively checks for further parts of the driving maneuver under consideration.
The recursion starts at the end of the previous part prolonged by a minimal gap time allowing also overlaps. If
all following parts for this driving maneuver under consideration are valid, the entire driving maneuver is
returned, otherwise the recursion is terminated.

The \gls{SFC}-based event detector is inspired by the design of the BF\_primitive query; however, the identification
of relevant entries is optimized. The \gls{SFC} query requires a search mask and, in contrast to BF, a
database index that is representing the single-dimensional, Morton-encoded spectrum; cf.~Sec.~\ref{sec:ZEBRA} for
details. For each search mask stage (ie., area-of-interest in this case), we determine the relevant Morton-codes to
match on the single-dimensional spectrum. After matching the relevant Morton-codes, the corresponding relevant
time points are retrieved and sorted by ascending order of time. The maneuver detection based on relevant time points
is then similar to the BF\_primitive query.

We evaluate efficiency by comparing the durations of the queries and efficacy by determining false/positives and
false/negatives. Since the database is not annotated, we consider the output of BF-primitive as ground truth as the 
process of BF-primitive is very similar to the intuition of experienced domain experts. We require and verify that all methods detect the same events. The dependent variables of
the experiment are (d1) duration of a query and (d2) false positive and false negative maneuvers. We report
the number of detected maneuvers as independent variable.

The controlled variables are (c1) the number of stages in a search mask, (c2) the dimensions of the search mask
stages (ie., bottom/left and top/right of the area-of-interest), and (c3) the size of the database.
Fixed parameters are the minimal and maximal duration of a driving maneuver ($T_{min\_dur} = 2,000 ms$,
$T_{max\_dur} = 3,000 ms$), the minimal and maximal gap between two valid driving maneuvers
($T_{min\_gap} = -200 ms$; $T_{max\_gap} = 2,000 ms$), and the maximum temporal distance between two valid
entries in case of outliers $T_{max\_outlier} = 50 ms$.

We conducted all experiments on a NUC 9 Pro computer powered by an Intel(R) Xeon(R) E-2286M
16-core CPU with 64GB RAM running Linux kernel 5.3.19-6 with 2x Sabrent Rocket Q (firmware RKT30Q.3)
8TB NVMe SSDs used with the ZFS filesystem in a striped data pool running ZFS version 2.5.1 and Linux
kernel module 2.4.1.

\begin{figure*}[t!]
    \centering
    \includegraphics[width=\textwidth]{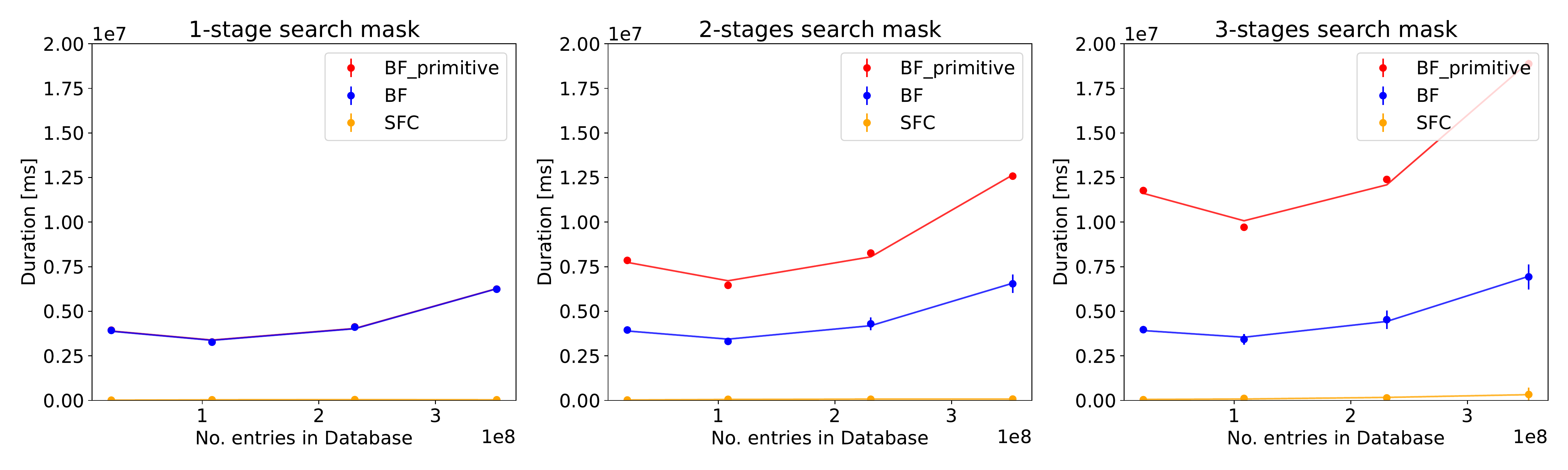}
    \caption{Query duration for varying search mask stages and database sizes. Please note that the red (BF\_primitive) and blue (BF) values overlap in the left diagram and hence, only the blue values are visible.}
    \label{fig:queryduration}
\end{figure*}

We consider search masks with one, two, or three stages and randomly generated five of these search masks each resulting in a
total of 15 search masks. We require that each search mask detects at least one valid maneuver in the overall
database that we verified with a \gls{SFC}-query. As a first step, we search for maneuvers for each search mask with
each query method (Bf\_primtive, BF, SFC) using the entire database (352,477,057 entries) resulting in a total of 45 queries. To investigate
the impact of database size on the query duration, we randomly extract three snippets with different sizes
(230,774,696 entries, 22,042,147 entries, 108,402,558 entries) from the overall database and apply all 45 queries
to each database excerpt respectively. The overall experiment took approx.~223h.%

We show the query durations separated by the number of search masks and arranged according to the database size
in Fig.~\ref{fig:queryduration}. From left to right, we plot results for the search masks with 1--3 stages. The
x-axis depicts the number of entries that are contained in the database or database excerpt, and the y-axis
describes the duration of the maneuver queries. A dot in the chart represents the mean of five
randomly generated search masks and a vertical line the spread of their minimum/maximum query duration. The
line connecting multiple dots shows a $2^\text{nd}$-order polynomial trend line.

The SFC query is clearly faster than the improved BF and BF-primitive queries.
The query duration increases for all approaches with growing database sizes and BF-primitive query is most and
SFC-query is least impacted by the database size as expected. Moreover, the number of stages significantly impacts
the BF-primitive query, while the improved BF mainly has similar query durations. The SFC-query outperforms both
query methods with respect to shortest query durations independently of search mask stages and database sizes.

\noindent\fbox{%
\begin{minipage}{.96\columnwidth}
\textbf{Answering RQ-2: What is the performance of event retrieval based on \glspl{SFC}?}\\
Our experiments showed that \gls{ZEBRA} identifies the same events as the na\"{i}ve brute force method
but significantly outperforms both brute force implementations in execution time. \gls{ZEBRA} provides
the possibility to efficiently query maneuvers from large-scale databases independently of search mask
complexity and dimensions of area-of-interests.
\end{minipage}
}\\

\subsection{Discussion}

Common to the illustrating application examples for the emergency braking and the lane change maneuvers as well
as the experimental setup as described before is that the transformation of the underlying data samples from
their multi-dimensional value space into the corresponding, single-dimensional spectrum must be conducted only
once \emph{irrespective} of the type of events to be spotted and how often any event detection queries shall be
executed. This is a strong benefit of the presented novel approach compared to rule-based event detection based
on annotations that in the worst case would need to re-process all data again on any change of the implementation.

In contrast to event detectors, which are directly applied to all entries of a data set and as a consequence
are not computationally efficient by design, event detectors operating on the single-dimensional spectrum are
not dependent on the amount of data to process due to its design operating on a fixed range to assess 
stripe ordering, temporal occurrence and spread of stripes, and stripe distribution. The only influencing
factor on the performance of \glspl{SFC}-based event detectors is the growing amount of entries in the lists with
candidate time points that are subsequently needed to evaluate temporal relations. Hence, we can conclude that
\gls{ZEBRA} is a computationally efficient approach for event detection.

Furthermore, the general architecture behind \gls{ZEBRA} correlates certain patterns from the single-dimensional
spectrum including temporal properties for stripe occurrence, spread, and distribution. Therefore, \gls{ZEBRA}
preserves the analytical explainability of event modeling and identification. As such, the presented approach of
\gls{SFC}-based event detectors can also be considered as a more suitable means for systematic event description
and identification in the context of, eg., safety-critical systems engineering, where systematic and traceable
safety argumentation is needed.

\section{Conclusion and Future Work} %
\label{sec:conclusion}

In this paper, we present a novel approach to model and retrieve events in multi-dimensional
data spaces called \gls{ZEBRA} by combining (a) dimensionality reduction of \glspl{SFC} resulting in
single-dimensional representations of values as \emph{Zebra-stripes}, with (b) the temporal properties for
stripe ordering, temporal occurrence and spread of stripes, and stripe distribution.
Computationally efficient event detection and retrieval is enabled when using the single-dimensional
values as additional index into key/value-databases. We illustrate \gls{ZEBRA} for maneuver detection in an
automotive data set and evaluate its performance on a large-scale automotive data set, where we could
reduce the processing time from initially approximately 69 minutes using an $O(n)$ event detector to
21 seconds using \gls{ZEBRA}.

The combination of dimensionality reduction of \glspl{SFC} with temporal properties from an
\gls{ODD} of a data space unlocks computationally efficient information retrieval. We foresee further
very relevant research and application directions such as real-time event classification, event similarity
identification in large data sets, or \gls{ODD}-agnostic event modeling for interactive data exploration
on large-scale data for improved edge-case modeling. The outlined \gls{SFC}-based approach is addressing
a relevant topic from industry that is even enabling further research opportunities for event modeling and
detection offering analytical explainability of detected events. Additionally, more research is needed to also 
integrate further unstructured data samples from video-, radar-, or lidar-sensors into the dimensionality
reduction processing step to allow for more detailed event detectors.

\section*{Acknowledgment}
This work has been partially been funded by the PhD program ``Responsible AI in the Digital Society'' funded by the Ministry for Science and Culture of Lower Saxony, Germany. This work has been partially supported by Vinnova, Grant Number 2021-05027.

\balance
\bibliographystyle{IEEEtran}
\bibliography{IEEEabrv,bib.bib}

\end{document}